\documentclass{PoS}

\usepackage{subcaption}

\newcommand{\hijingpp}{\texttt{HIJING++}~}
\newcommand{\pythia}{\texttt{Pythia8}~}
\newcommand{\fortran}{\texttt{FORTRAN}~}
\newcommand{\rivet}{\texttt{RIVET}~}
\newcommand{\cernroot}{\texttt{ROOT}~}
\newcommand{\hepmc}{\texttt{HepMC}~}

\title{Introducing HIJING++: the Heavy Ion Monte Carlo Generator for the High-Luminosity LHC Era}

\ShortTitle{Introducing HIJING++}

        \author{\speaker{G\'abor B\'ir\'o}$^{ab}$, Gergely G\'abor Barnaf\"oldi$^{b}$, G\'abor Papp$^{a}$, Miklos Gyulassy$^{bcde}$, P\'eter L\'evai$^{b}$, Xin-Nian Wang$^{cd}$ and Ben-Wei Zhang$^{c}$\\
        \llap{$^{a}$}
        Institute for Physics, E\"otv\"os Lor\'and University\\
        1/A P\'azm\'any P. S\'et\'any, H-1117, Budapest, Hungary\\
        \llap{$^{b}$}
        Wigner Research Centre for Physics of the Hungarian Academy of Sciences\\
        29-33 Konkoly-Thege Mikl\'os Str, H-1121 Budapest, Hungary\\
        \llap{$^{c}$}
        Key Laboratory of Quark \& Lepton Physics (MOE) and Institute of Particle Physics\\
        Central China Normal University, Wuhan 430079, China\\
        \llap{$^{d}$}
        Nuclear Science Division, MS 70R0319, Lawrence Berkeley National Laboratory\\
        Berkeley, California 94720 USA \\
        \llap{$^{e}$}
        Pupin Lab MS-5202, Department of Physics, Columbia University\\
        New York, NY 10027, USA\\
        E-mail: \email{biro.gabor@wigner.mta.hu}, \email{barnafoldi.gergely@wigner.mta.hu}, \email{pg@ludens.elte.hu}
}

\abstract{
Beyond 2025 we will enter the High-Luminosity era of the LHC, right after the upgrades of the third Long Shutdown of the Large Hadron Collider (LHC). The ongoing state-of-the-art experimental instrument upgrades require high-performance simulation support in the background, that is modern, robust, and comes with long term support. The original \fortran based \texttt{HIJING} (Heavy Ion Jet INteraction Generator) has been used intensively since almost three decades by the high energy physics and heavy-ion community. However, it is getting overly challenging to conform to these new requirements. Our novel Monte Carlo event generator, the \hijingpp is the successor of the old \fortran version containing all the physics that its predecessor have. Moreover, among others a flexible module handling layer and an analysis interface is also introduced. This latter supports the most popular event container formats such as simple \texttt{ascii}, \cernroot and the \hepmc format, together with \rivet support. In this paper we compare the pre-release results of \hijingpp with proton-proton experimental data and \pythia calculations.
}


\FullConference{International Conference on Hard and Electromagnetic Probes of High-Energy Nuclear Collisions\\
30 September - 5 October 2018\\
		Aix-Les-Bains, Savoie, France}

\begin{document}

\section{Introduction}

The main motivation during the development of \hijingpp~\cite{HIJING,HIJING2, HPP1,HPP2,HQP} is to obtain an event generator of high-energy proton-proton and heavy-ion collisions that is easily maintainable, extendible and works effectively with high throughput. To achieve this, we built up \hijingpp from scratch and performed an initial tune of the parameters for RHIC and LHC energies. In this way it was possible to design a framework that is powerful not just computationally, but thanks to modern C++ techniques also has an effective analysis interface with multiple compatible formats. Among a simple \texttt{ascii} and the popular \cernroot\cite{ROOT} formats the interface incorporates also the \hepmc\cite{HEPMC} event format which is a \textit{de facto} standard in High Energy Physics. In this contribution we present the first consistent results for pseudorapidity and transverse momentum distribution at LHC energies using \rivet\cite{RIVET} analyses.

\section{Run settings}

The main input parameters for \hijingpp can be listed in command cards familiar from \pythia. These include the kinematical parameters like collision energy, the type of the target and projectile (proton-proton in this contribution) and the PDF sets to be used. For this latter during the ongoing tuning we have chosen the \texttt{nCTEQ15} sets that are available in the \texttt{LHAPDF6} framework~\cite{LHAPDF, NCTEQ15}. Our argument for these (nuclear) PDF sets is mainly practical: it is a comprehensive parametrization for all common heavy and light nuclei including the protons as well, where we don't expect any nuclear effects to play role. We checked this assumption also with other, purely proton PDFs like \texttt{CT14nlo}~\cite{CT14NLO}. We have found that the \hijingpp results using the two different PDF sets are consistent, therefore we decided to use the one that has extensions also for heavier nuclei.
For the \pythia\cite{PYTHIA} comparisons we used the \texttt{Monash 2013} tune~\cite{MONASH}.

The pre-release version of \hijingpp is already capable to reproduce a variety of experimental results without any further artificial post-tuning. The \hepmc event format also provides a possibility to use \rivet for analyzing the output in a standardized way, fine tune the physically irrelevant numerical parameters of the Monte Carlo event generator and validate the results more effectively. In the following section we present the \rivet results based on ALICE publications \cite{ALICE1, ALICE2, ALICE3}.

\section{Results}


\subsection{Pseudorapidity distributions}

\begin{figure}[h!]
    \centering
    \begin{subfigure}[t]{0.329\textwidth}
        \includegraphics[width=\textwidth]{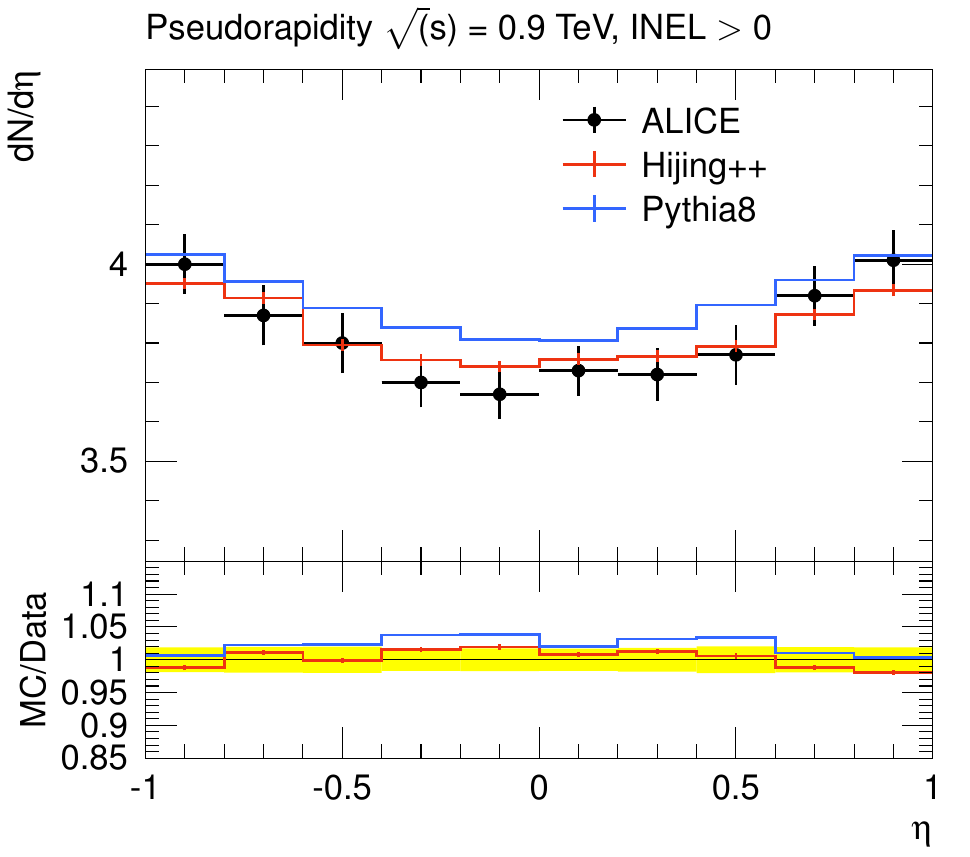}
        \caption{}
        \label{dndeta1}
    \end{subfigure}
    \begin{subfigure}[t]{0.329\textwidth}
        \includegraphics[width=\textwidth]{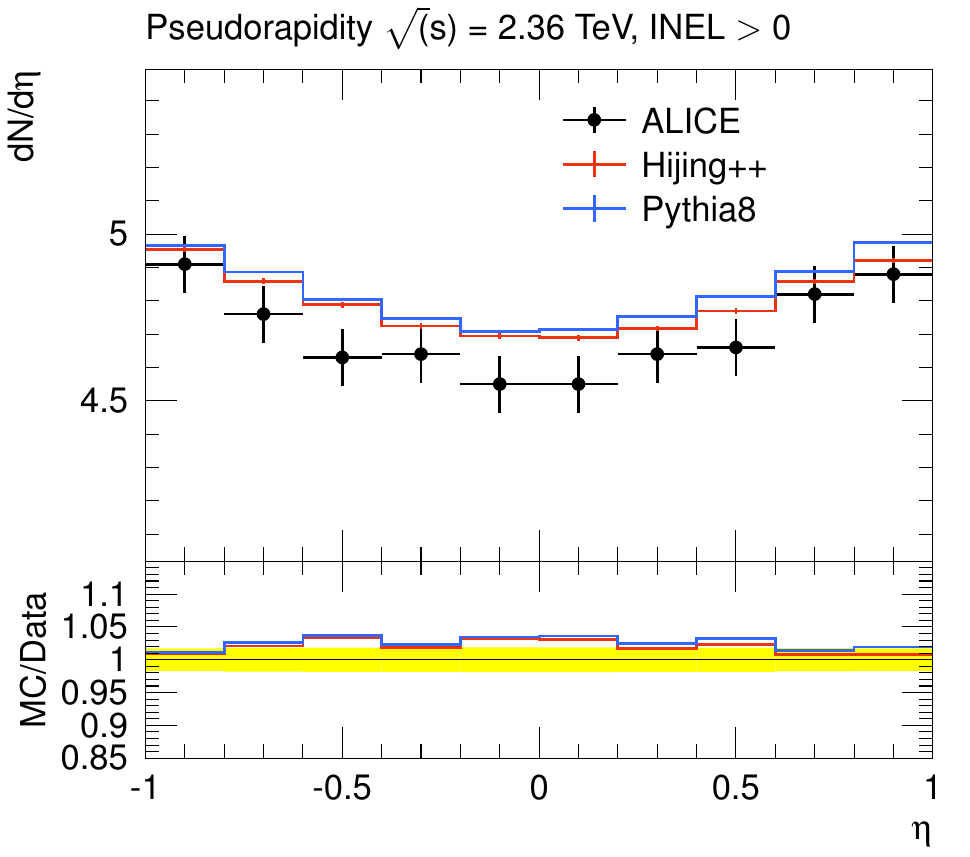}
        \caption{}
        \label{dndeta2}
    \end{subfigure}
    \begin{subfigure}[t]{0.329\textwidth}
        \includegraphics[width=\textwidth]{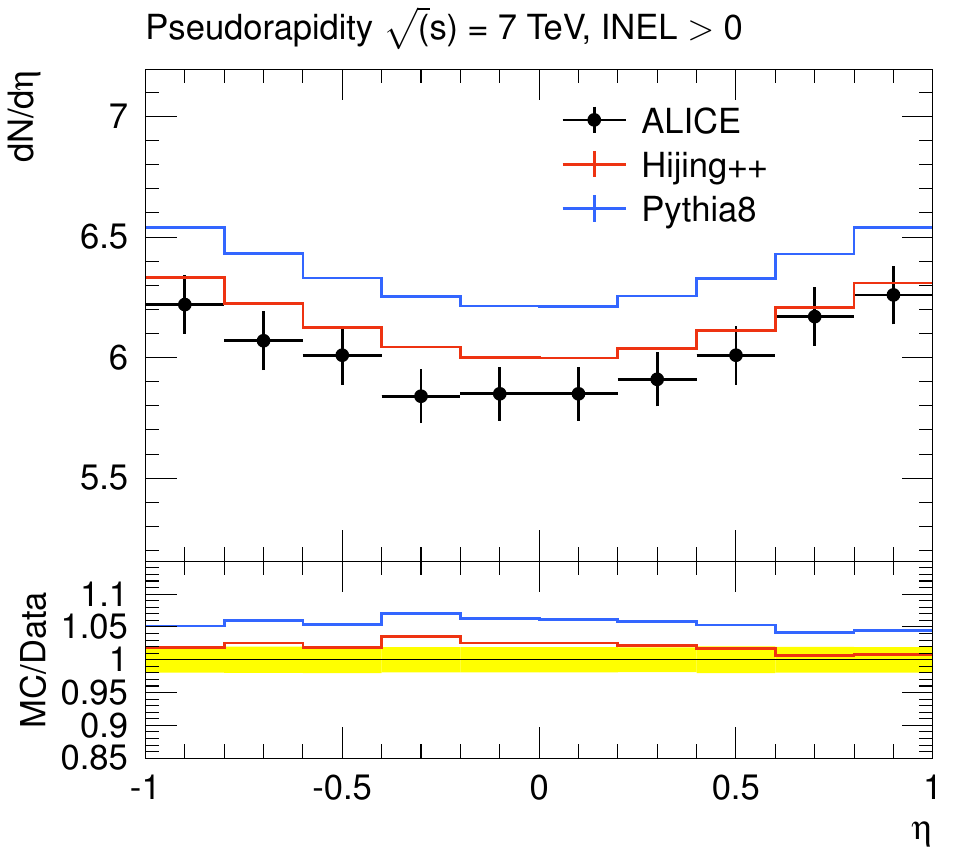}
        \caption{}
        \label{dndeta3}
    \end{subfigure}
    \caption{Pseudorapidity distributions of charged hadrons at $\sqrt{s}=0.9$ TeV (\textit{left} panel), 2.36 TeV (\textit{middle} panel) and 7 TeV (\textit{right} panel) in proton-proton collisions~\cite{ALICE1}.}
    \label{dndeta}
\end{figure}

On Figure \ref{dndeta} the pseudorapidity distributions of charged hadrons in proton-proton collisions at $\sqrt{s}=0.9$ TeV, $2.36$ TeV and $7$ TeV are plotted (top \textit{left, middle} and \textit{right panel} respectively)~\cite{ALICE1}. On the bottom panels the Monte Carlo over data ratios can be seen. At the lowest and highest collision energies the current pre-release \hijingpp results describe better the experimental multiplicity than \pythia with the default settings, while at $\sqrt{s}=2.36$ TeV the results of the two event generators are almost the same. Although the current \hijingpp is still the pre-release version with the tuning is currently ongoing, the pseudorapidity distributions are already compatible with the experimental results within $\sim 4\%$ precision.

\subsection{Identified hadron $p_T$ spectra}

On Figure \ref{fig:spectra2} the $p_T$ distributions of identified $\pi^0$ (\textit{left panels}) and $\eta$ (\textit{right panels}) particles measured at midrapidity are plotted at $\sqrt{s}=2.76$ TeV and $7$ TeV center-of-mass energy and compared with ALICE experimental results~\cite{ALICE2, ALICE3}. The Monte Carlo over data ratios are plotted on the bottom panels as well.

\begin{figure}[h!]
    \centering
    \begin{subfigure}[t]{0.46\textwidth}
        \includegraphics[width=\textwidth]{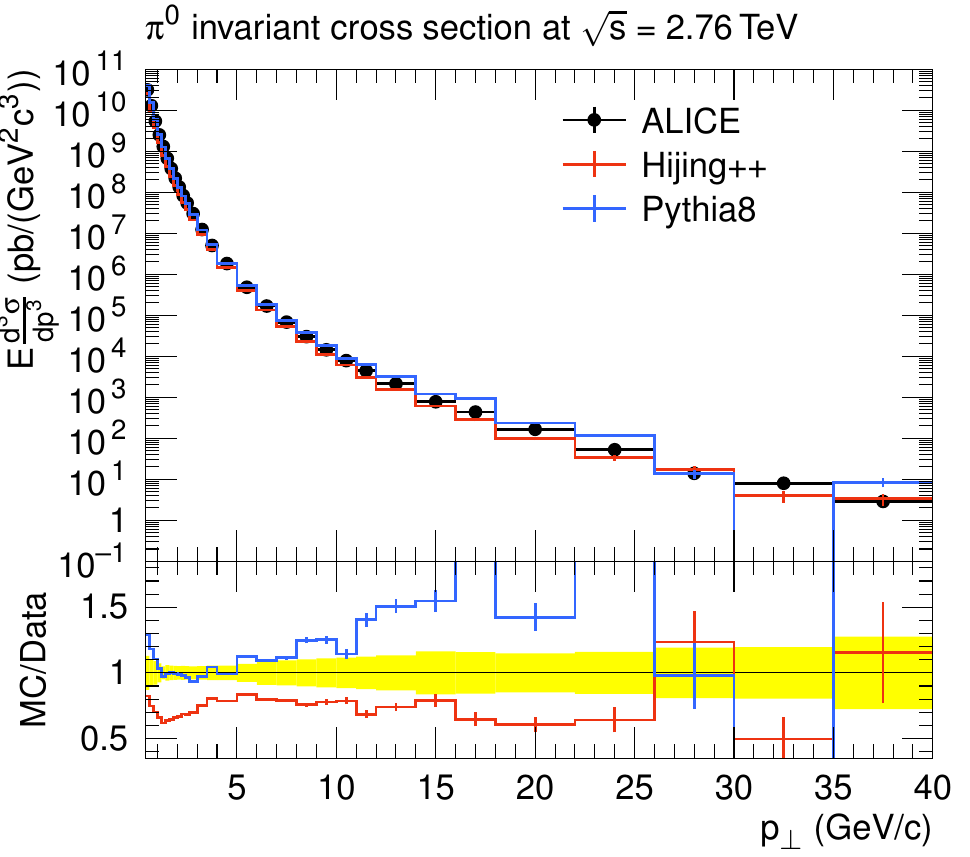}
        \caption{}
        \label{fig:spectra1}
    \end{subfigure}
    \begin{subfigure}[t]{0.46\textwidth}
        \includegraphics[width=\textwidth]{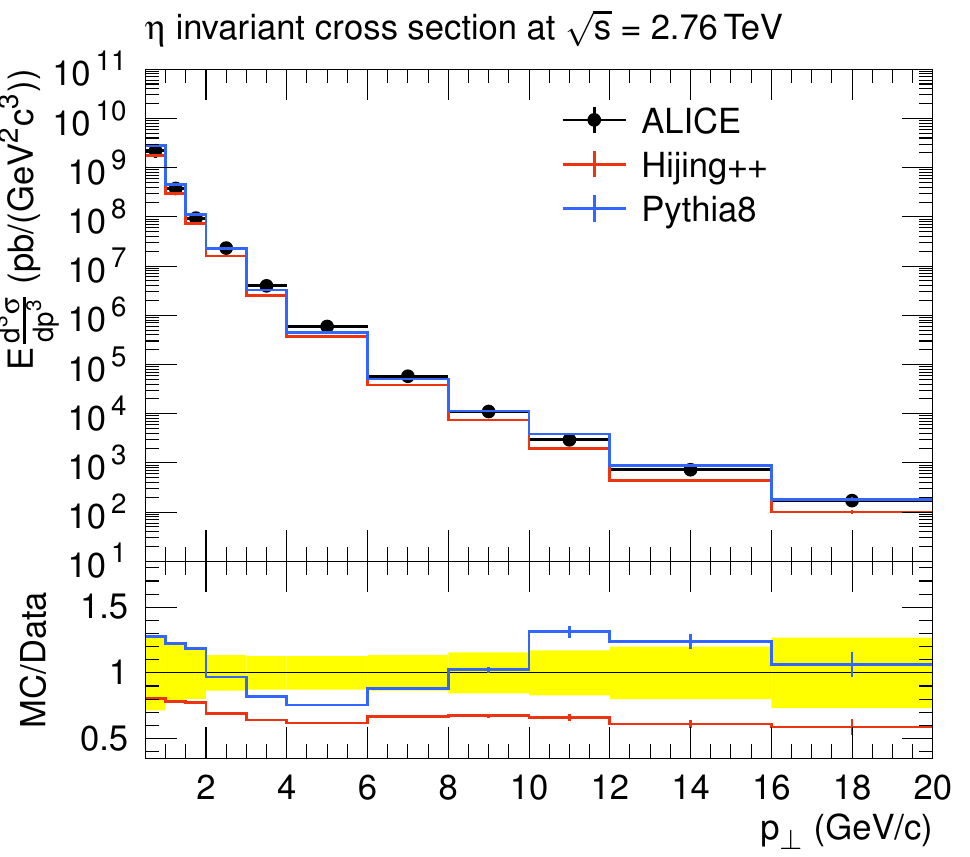}
        \caption{}
        \label{fig:spectra2}
    \end{subfigure}
    \begin{subfigure}[t]{0.46\textwidth}
        \includegraphics[width=\textwidth]{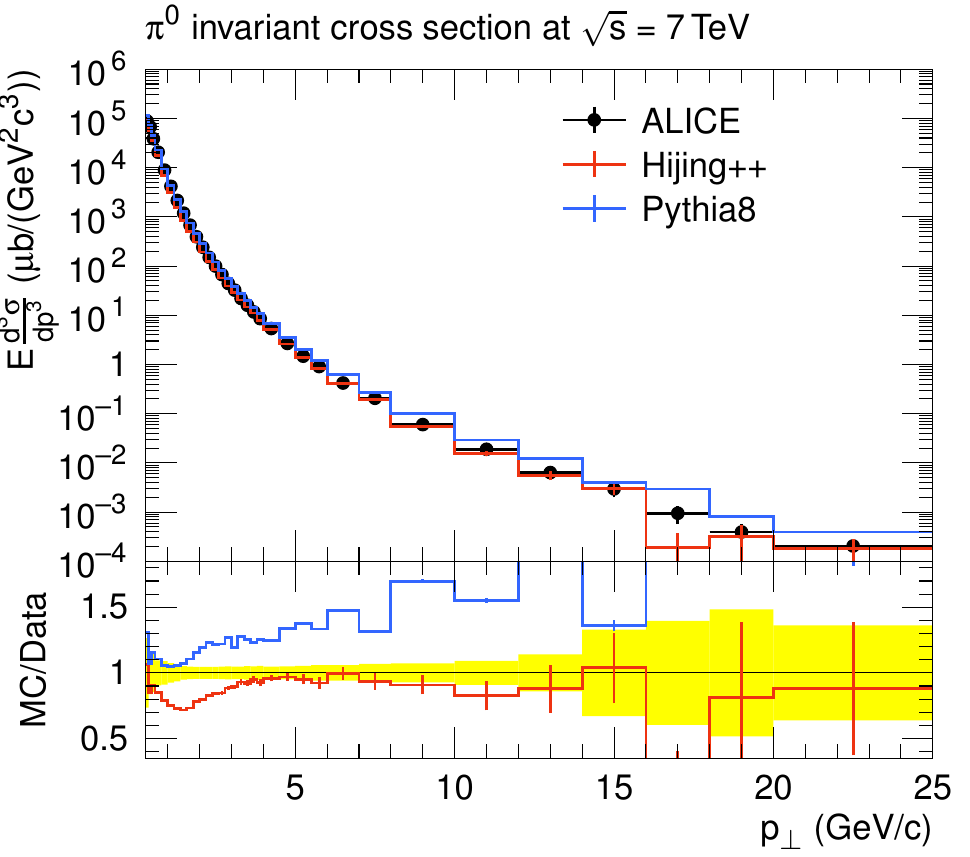}
        \caption{}
        \label{fig:spectra4}
    \end{subfigure}
    \begin{subfigure}[t]{0.46\textwidth}
        \includegraphics[width=\textwidth]{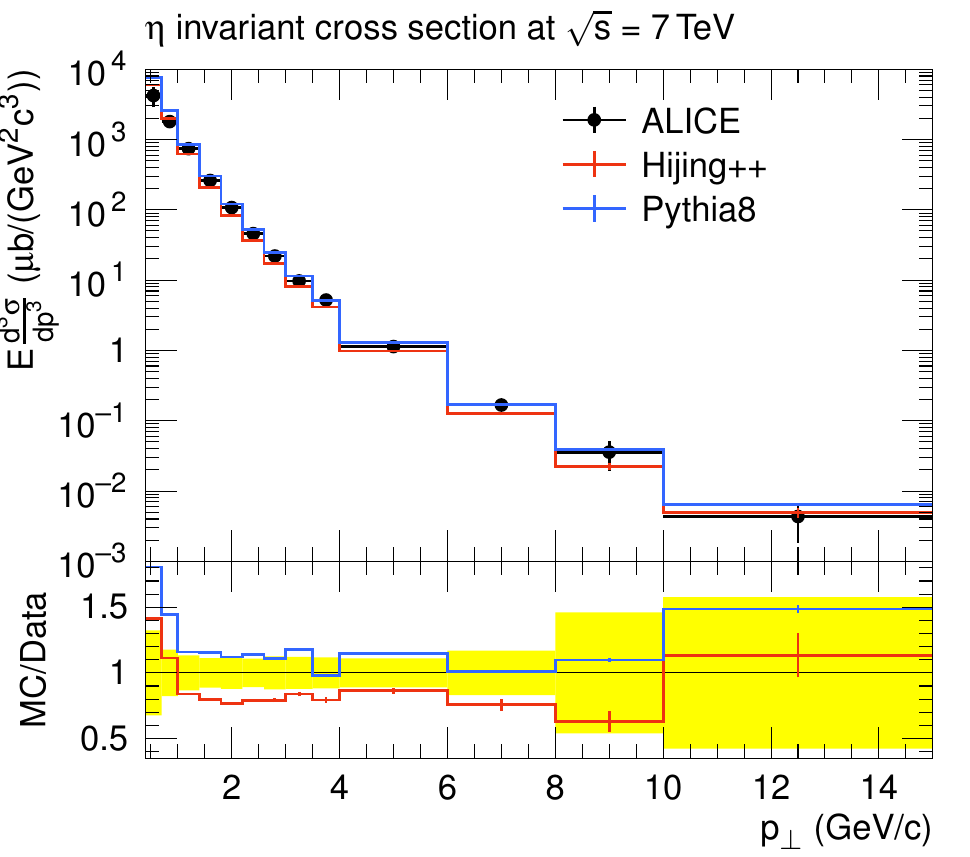}
        \caption{}
        \label{fig:spectra5}
    \end{subfigure}
    \caption{Spectra of identified $\pi^0$ (\textit{left} panel) and $\eta$ (\textit{right} panel) particles at $\sqrt{s}=2.76$ TeV (\textit{top} panels) and $7$ TeV  (\textit{bottom} panels) in proton-proton collisions~\cite{ALICE2, ALICE3}.}
    \label{fig:spectra2}
\end{figure}

\noindent\hijingpp is performing better overall than \pythia describing the yield of the $\pi^0$ particles. Although in the low-$p_T$ regime \hijingpp slightly underestimating the yield, at higher $p_T$ (i.e. $p_T \gtrsim 4$ GeV/c) the agreement is within 10\%, especially at the higher $\sqrt{s}=7$ TeV energy. For the $\eta$ particles the \pythia calculations are closer to the experimental measurements, the \hijingpp underestimates the yield. The agreement is the best at the mid-$p_T$ region at $\sqrt{s}=7$ TeV, where the \hijingpp results lie within $\sim10\%$ of the experimental data.

\begin{figure}[h!]
    \centering
    \begin{subfigure}[t]{0.46\textwidth}
        \includegraphics[width=\textwidth]{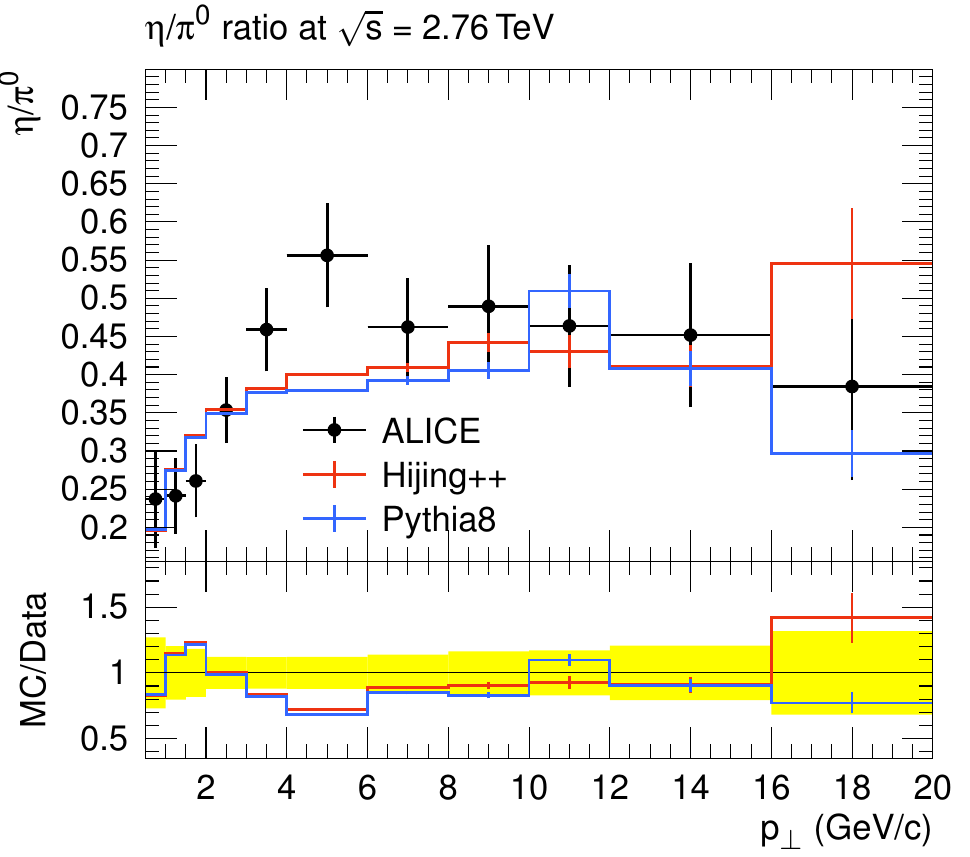}
        \caption{}
        \label{fig:spectra3}
    \end{subfigure}
    \begin{subfigure}[t]{0.46\textwidth}
        \includegraphics[width=\textwidth]{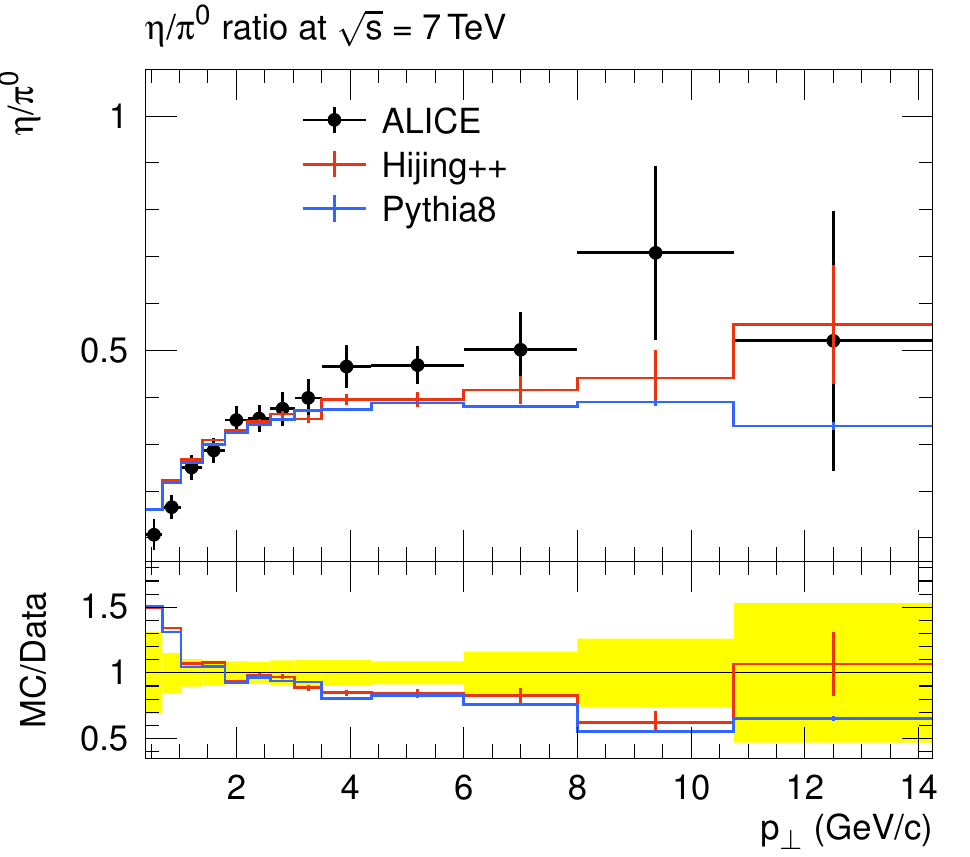}
        \caption{}
        \label{fig:spectra6}
    \end{subfigure}
    \caption{Ratio of identified $\pi^0$ and $\eta$ particles at $\sqrt{s}=2.76$ TeV and $\sqrt{s}=7$ TeV proton-proton collisions~\cite{ALICE2, ALICE3}.}
    \label{fig:ratios}
\end{figure}

On Figures \ref{fig:ratios} the ratio of the identified $\eta$ and $\pi^0$ particles measured at midrapidity at $\sqrt{s}=2.76$ TeV and $\sqrt{s}=7$ TeV center-of-mass energies are shown along with the Monte Carlo over data ratios.

\noindent Despite the differences between the identified $\pi^0$ and $\eta$ yields of \hijingpp and \pythia, the ratios show strong similarities both in $\sqrt{s}=2.76$ and 7 TeV energies. For both event generators the agreement with the experimental measurements is well within $\sim10\%$ at the low- and high-$p_T$ region at $\sqrt{s}=2.76$ TeV, while at the mid-$p_T$ around $p_T=5$ GeV/c the theoretical calculations are flatter than the ALICE results. At $\sqrt{s}=7$ TeV the experimental points going to the 0.5 value faster, the increasing tendency of the Monte Carlo results is slower. The best agreement is around $p_T\sim3$ GeV/c, where the data-MC disagreement is only a few \%.

\section{Summary and outlook}
 
In this paper we presented the theoretical calculations of the pre-release version of the new \hijingpp Monte Carlo particle event generator. We compared the pseudorapidity and $p_T$ distributions of charged and identified hadrons with experimental proton-proton measurements and \pythia results using the \hepmc format and \rivet analyses. We showed that although the fine-tuning of \hijingpp is currently ongoing, the current version already has a good agreement with experimental proton-proton data. The present calculations show, that despite the $p_T$ spectra of identified $\pi^0$ and $\eta$ hadrons differ substantially in \hijingpp and \pythia, their ratios are well in agreement both at $\sqrt{s} = 2.76$ TeV and 7 TeV center-of-mass energies.



\begin{thebibliography}{99}
    \bibitem{HIJING} 
    X.N. Wang, M. Gyulassy, 
    {\em Phys. Rev.} {\bf D44} (1991) 3501.
    
    \bibitem{HIJING2} 
    W.T. Deng, X.N. Wang, R. Xu, 
    {\em Phys. Rev.} {\bf C83} (2011) 014915.

    \bibitem{HPP1} 
    G.G. Barnaf\"oldi, G. B\'ir\'o, M. Gyulassy, S.M. Haranoz\'o, P. L\'evai, G. Ma, G. Papp, X.N.Wang, B.W. Zhang, 
    {\em Nucl. and Part. Phys. Proc.} {\bf 289-290} (2017) 373-376.
    
    \bibitem{HPP2} 
    G. Papp, G.G. Barnaf\"oldi, G. B\'ir\'o, M. Gyulassy, Sz.M.Harangoz\'o, G. Ma, P. L\'evai, X.N. Wang, B.W. Zhang, 
    {\em Accepted to P.o.S.} (2018) arXiv:1805.02635
    
    \bibitem{HQP} 
    G. B\'ir\'o, G. Papp, G.G. Barnaf\"oldi, D. Nagy, M. Gyulassy, P. L\'evai, X.N. Wang, B.W. Zhang, 
    {\em Accepted to MDPI Proceedings} (2018) arXiv:1811.02131
    
\bibitem{ROOT} https://root.cern.ch/ (25.12.2018.)

\bibitem{HEPMC}
  M.~Dobbs and J.~B.~Hansen,
  Comput.\ Phys.\ Commun.\  {\bf 134} (2001) 41.
  
\bibitem{RIVET} A.~Buckley, J.~Butterworth, L.~Lonnblad, D.~Grellscheid, H.~Hoeth, J.~Monk, H.~Schulz and F.~Siegert,
  {\em Comput.\ Phys.\ Commun.\ } {\bf 184} (2013) 2803
  doi:10.1016/j.cpc.2013.05.021
  [arXiv:1003.0694 [hep-ph]].
  
  \bibitem{LHAPDF} A. Buckley, J. Ferrando, S. Lloyd, K. Nordstr\"om, B. Page, M. R\"ufenacht, M. Sch\"onherr, G. Watt,  {\em Eur. Phys. J.} {\bf 2015}, {\em C75} 3, 132
  
  \bibitem{NCTEQ15}
    K.~Kovarik {\it et al.},
    Phys.\ Rev.\ D {\bf 93} (2016) no.8,  085037
    [arXiv:1509.00792 [hep-ph]].

\bibitem{CT14NLO}
  S.~Dulat {\it et al.},
  Phys.\ Rev.\ D {\bf 93} (2016) no.3,  033006
  [arXiv:1506.07443 [hep-ph]].

  \bibitem{PYTHIA} 
  T. Sj\"ostrand, 
  {\em Comput. Phys. Commun.} {\bf 191} (2015) 159.
  
  \bibitem{MONASH}
  P.~Skands, S.~Carrazza and J.~Rojo,
  Eur.\ Phys.\ J.\ C {\bf 74} (2014) no.8,  3024
  [arXiv:1404.5630 [hep-ph]].

\bibitem{ALICE1}
  K.~Aamodt {\it et al.} [ALICE Collaboration],
  Eur.\ Phys.\ J.\ C {\bf 68} (2010) 345
  [arXiv:1004.3514 [hep-ex]].

\bibitem{ALICE2}
S.~Acharya {\it et al.} [ALICE Collaboration],
Eur.\ Phys.\ J.\ C {\bf 77} (2017) no.5,  339
 [Eur.\ Phys.\ J.\ C {\bf 77} (2017) no.9,  586]
[arXiv:1702.00917 [hep-ex]].
  
\bibitem{ALICE3}
  B.~Abelev {\it et al.} [ALICE Collaboration],
  Phys.\ Lett.\ B {\bf 717} (2012) 162
  [arXiv:1205.5724 [hep-ex]].

\bibitem{GSL} 
M. Galassi et al, 
{\em GNU Scientific Library Reference Manual} (3rd Ed.), ISBN 0954612078

\bibitem{VEGAS} 
G.P. Lepage, 
{\em J. of Comp. Phys.} {\bf 27} (1978) 192-203.





\end{thebibliography}
\end{document}